\documentclass[amsmath,pra,twocolumn,showpacs]{revtex4}
\usepackage{graphics}

\begin{document}     

\title{Frequency modulation and pulse compression by coherent multimode   molecular motion}
\author{Fam Le Kien} 
\altaffiliation{On leave from Department of Physics, University of
Hanoi, Hanoi, Vietnam; also at Institute of Physics, National Center for 
Natural Sciences and Technology, Hanoi, Vietnam} 
\affiliation{Department of Applied Physics and Chemistry, 
University of Electro-Communications, Chofu, Tokyo 182-8585, Japan\\
CREST, Japan Science and Technology Corporation (JST), Chofu, Tokyo 182-8585, Japan}
\author{K. Hakuta}
\affiliation{Department of Applied Physics and Chemistry, 
University of Electro-Communications, Chofu, Tokyo 182-8585, Japan\\
CREST, Japan Science and Technology Corporation (JST), Chofu, Tokyo 182-8585, Japan}
\author{A. V. Sokolov}
\affiliation{Department of Physics and Institute for Quantum Studies,
Texas A\&M University, College Station, TX 77843-4242}
\date{\today}

\begin{abstract}
We study beating of a probe field with a  time-varying susceptibility in
a coherently prepared Raman medium. We consider the 
general case of an arbitrary variation of  susceptibility, which 
corresponds to a superposition of an arbitrary number of excited Raman 
transitions.
We derive a general analytical  solution and conservation relations for this process. We show that the interference between  Raman polarizations may  substantially affect  frequency modulation and pulse compression for the probe field.
\end{abstract}

\pacs{42.65.Re, 32.80.Qk, 42.50.Gy, 42.65.Sf}
\maketitle

We have recently presented a detailed analysis of ultrashort laser pulse 
compression by parametric beating with a sinusoidal molecular oscillation 
\cite{Kien02}.  In this Brief Report, we extend the previous work to the 
general case of an arbitrarily complex molecular motion. We assume a wave-like 
molecular excitation, produced by external fields in a dispersionless 
medium.  Molecular motion results in a time-varying susceptibility, which 
(in the absence of dispersion) leads to time variations in phase and group 
velocities for a probe field  \cite{Kien02}.  
We analyze the effect of these variations on pulse propagation, derive an analytical solution in terms of an 
integral  for a time transformation, and use it to obtain 
conservation relations.

It was shown recently that coherently driven molecular oscillations can 
produce frequency modulation with ultrabroad bandwidth 
\cite{Modulation,all experiments}, and result in subfemtosecond \cite{some theory,Sokolov01,korn02} 
and subcycle  \cite{HarrisSeries} pulse compression.  
A molecular oscillation can be either sinusoidal
(a single coherently excited Raman transition) \cite{Modulation,all experiments,some theory,Sokolov01,korn02}  
or more complex (multimode vibrational or rotational 
molecular wave packets, which correspond to a coherent phased excitation of 
many Raman transitions) \cite{Ivanov02, Kapteyn01}.   
Molecular wave packets can be excited either 
impulsively \cite{korn02, Ivanov02, Kapteyn01}, or by applying several 
quasi-monochromatic laser fields \cite{HarrisSeries}.  
Generalization of our previous results \cite{Kien02} to the important 
case of a complex molecular motion will provide  insight into  the Raman pulse
compression techniques. 

The essence of our result is simple:  A pulse is  compressed or stretched 
by time-varying phase and group velocities, such that envelope 
compression or stretching is accompanied by frequency increase or decrease, respectively.  
Pulse amplitude and oscillation frequency are changed by reciprocal factors, such that the 
photon number, the pulse area, and the number of optical oscillations are 
conserved, while the pulse energy is not.  When the resultant change in 
the pulse bandwidth is small compared to the optical carrier frequency, 
pulse deformation by the complex non-sinusoidal molecular motion can be obtained as 
a multiplication of the effects of individual harmonic 
components of the motion.  
However, for stronger modulation this intuitive 
thinking can not be applied.  In order to emphasize this point, in the 
second part of our Report we consider an example of a biharmonic 
molecular excitation and show that frequency modulation and pulse compression are significantly 
affected by the interference between two Raman polarizations.

We study the propagation of a  probe field $E$ in a Raman medium, characterized by a polarization $P$.  
In the local time $\tau=t-z/c$, the reduced wave equation reads 
\begin{equation}
\frac{\partial E }{\partial z}  = 
-\frac{1}{2\epsilon_0 c} \frac{\partial { P}}{\partial \tau}. 
\label{1}
\end{equation} 
We assume that dispersion is negligible. 
The instantaneous  susceptibility  is defined as 
$\chi=P/\epsilon_0 E$.
We assume that 
$\chi=\chi_0+\chi_m(\tau-z/v)$,
where $\chi_0$ is the linear time-independent susceptibility, $\chi_m$ is
the Raman time-varying susceptibility, and $v=2c/\chi_0$ is the average 
phase and group velocity in the local time coordinates.

We use the reduced local time $\eta=\tau-z/v$. In the coordinates
$z$ and $\eta$, Eq. (\ref{1}) becomes
\begin{equation}
\frac{\partial  }{\partial z}E(z,\eta)  = 
-\frac{1}{2 c} \frac{\partial {}}{\partial \eta}\chi_m(\eta) E(z,\eta). 
\label{4}
\end{equation} 
The solution to Eq. (\ref{4}) is found to be
\begin{equation}
 E (z,\eta)= E _{\mathrm{in}}(s) G(\eta) , 
\label{5}
\end{equation}
where $E _{\mathrm{in}}(s)= E (z=0,s)$ is the input field,  
$s$  is the input time  determined from the output time $\eta$ by   
\begin{equation}
\int_{s}^{\eta}\frac{d\theta}{\chi_m(\theta)}=\frac{z}{2c},
\label{6}  
\end{equation}
and  $G(\eta)$ is the compression factor  given by  
\begin{equation}
G(\eta) = 
\frac{\chi_m(s)}{\chi_m(\eta)}.
\label{7}
\end{equation} 
Differentiating Eq. (\ref{6}) with respect to $\eta$, we obtain 
\begin{equation}
\frac{ds}{d\eta}=G(\eta).
\label{8}
\end{equation} 
According to Eqs. (\ref{5}) and (\ref{8}), 
the height and    duration of optical oscillations in the probe field  
are changed by the reciprocal factors $E(z,\eta)/E_{\mathrm{in}}(s)=G(\eta)$ and 
$d\eta/ds=1/G(\eta)$, respectively.  
The value $G(\eta)>1$ ($G(\eta)<1$) indicates pulse compression (stretching) in the vicinity of $\eta$.
From Eqs. (\ref{5}) and (\ref{8}), we find  the relation
\begin{equation}
\int_{\eta_1}^{\eta_2} E(z,\eta)\, d\eta=\int _{s_1}^{s_2} E_{\mathrm{in}}(s)\, ds, 
\label{11}
\end{equation}
which describes the conservation of the pulse area.

We find from Eq. (\ref{8}) that the instantaneous oscillation frequency of the output field is   
\begin{equation}
\omega_{\mathrm{osc}}(\eta)=G(\eta)\omega_0,
\label{10}
\end{equation} 
where $\omega_0$ is the input frequency. 
As seen,  $\omega_{\mathrm{osc}}(\eta)$ 
is modulated in time by the compression factor $G(\eta)$. 
Using  Eqs. (\ref{5}), (\ref{8}), and (\ref{10}), we find 
\begin{equation}
\frac{c\epsilon_0}{2}\int_{\eta_1}^{\eta_2} \frac{E^2(z,\eta)}{\hbar\omega_{\mathrm{osc}}(\eta)}\, d\eta=
\frac{c\epsilon_0}{2}\int _{s_1}^{s_2} \frac{E_{\mathrm{in}}^2(s)}{\hbar\omega_0}\, ds .
\label{12}
\end{equation}
This relation describes the conservation 
of the photon number, which is always satisfied for Raman processes.

We introduce the mean  frequency  
$\bar \omega(\eta_1,\eta_2)=(\eta_2-\eta_1)^{-1}
\int _{\eta_1}^{\eta_2}  \omega_{\mathrm{osc}}(\eta) \, d\eta$ 
for the oscillations  in the time interval $(\eta_1,\eta_2)$.  
Then, we find from Eqs. (\ref{8}) and (\ref{10}) that
$\bar \omega(\eta_1,\eta_2)(\eta_2-\eta_1)= \omega_0(s_2-s_1)$,
that is,  the product of the pulse length and the mean frequency is constant during the 
propagation process. 
Furthermore,  Eq. (\ref{5}) says that 
$E(z,\eta)=0$ if $E_{\mathrm{in}}(s)=0$ and vice versa, that is, 
a zero of the input field at an input time $s$ leads to a zero of the output field at the corresponding output
time $\eta$. 
Hence, the number of optical oscillations is conserved. 

To get insight into the behavior of the compression factor $G$, 
we derive an explicit, approximate expression for this factor. 
For this purpose, we consider a particular case where the deviation of  an output time $\eta_i$  from     
its corresponding input time $s_i$ 
is small compared to the characteristic variation time $T_m$ of $\chi_m$, that is, $|s_i-\eta_i|\ll T_m$. 
We take   $\eta$  close to  $\eta_i$, and use the approximation
$\chi_m(\theta)=\chi_m(\eta_i)+\chi_m'(\eta_i)(\theta-\eta_i)$
to calculate 
the integral in Eq. (\ref{6}).  Then, we  find  
\begin{equation}
s-\eta_i+\frac{\chi_m(\eta_i)}{\chi_m'(\eta_i)}=\left[\eta-\eta_i+\frac{\chi_m(\eta_i)}{\chi_m'(\eta_i)}\right]
\exp\left[-\frac{z}{2c}\chi_m'(\eta_i) \right].
\label{18}
\end{equation}
Hence,   Eq. (\ref{8})  yields  
\begin{equation}
G(\eta_i)=\left. \frac{ds}{d\eta}\right|_{\eta=\eta_i}=\exp\left[-\frac{z}{2c}\chi_m'(\eta_i) \right].
\label{19}
\end{equation} 
Equation (\ref{19}) shows  that the  factor $G$ is approximately determined by the time derivative 
of the Raman susceptibility $\chi_m$, multiplied with the propagation length $z$.  
Note that,  around the chosen time $\eta_i$, the factor $G$ is multiplicative, that is, 
pulse deformation by the complex molecular motion can be obtained  as a multiplication of the  effects of individual  components of the motion.   
The assumption  $|s_i-\eta_i|\ll T_m$ requires 
\begin{equation}
\left|\frac{\exp\left[-(z/2c)\chi_m'(\eta_i) \right]-1}{\chi_m'(\eta_i)}\right|
\ll \frac{T_m}{|\chi_m(\eta_i)|}.
\label{20}
\end{equation}
Condition (\ref{20}) is satisfied when  
$(z/c)|\chi_m'(\eta_i)| \ll 1$ and  $(z/c) |\chi_m(\eta_i)| \ll T_m $, i.e.,  
when the Raman susceptibility and its modulation are small.  
In this case, we have     
$G(\eta_i)\cong 1-(z/2c)\chi_m'(\eta_i) \cong 1$,  indicating that  
the pulse compression   is weak. 
Condition (\ref{20}) is also satisfied  
when $\eta_i$ is a zero of $\chi_m$.    

In  general, $G$ is not multiplicative with respect to the individual susceptibilities.
To show this, we  consider the case where the Raman susceptibility $\chi_m$ is a sum of two sinusoids  corresponding to two individual Raman transitions, namely,   
$\chi_m=\chi_a+\chi_b$,
where
$\chi_{j}=\chi_{j}^{(0)}\sin(\omega_{j}\eta+\varphi_{j})$
for ${j}=a,b$.  
The  compression factor  corresponding to the individual component $\chi_{j}$
is given by the periodic comb function \cite{Kalosha and Kien,Kien02}
\begin{equation}
G_{j} = \frac{1}{e^{\alpha_{j} z} \cos^2 \frac{\omega_{j}\eta+\varphi_{j}}{2} 
+ e^{-\alpha_{j} z} \sin^2 \frac{\omega_{j}\eta+\varphi_{j}}{2}} ,  
\label{25}
\end{equation} 
where $\alpha_{j} =(\omega_{j}/2c) \chi_{j}^{(0)}$.
In general, the total compression factor $G$  is
different from the product of  the individual compression factors $G_a$ and $G_b$. The reason is that
the compression factor $G$ is produced by 
the time-varying Raman susceptibility $\chi_m$ via a nonlinear mechanism, see Eqs. (\ref{6}) and (\ref{7}). 
Due to the nonlinearity of this mechanism, the  individual susceptibility  components 
may interfere with each other  in beating with the  field.

\begin{figure}
\begin{center}
  \includegraphics{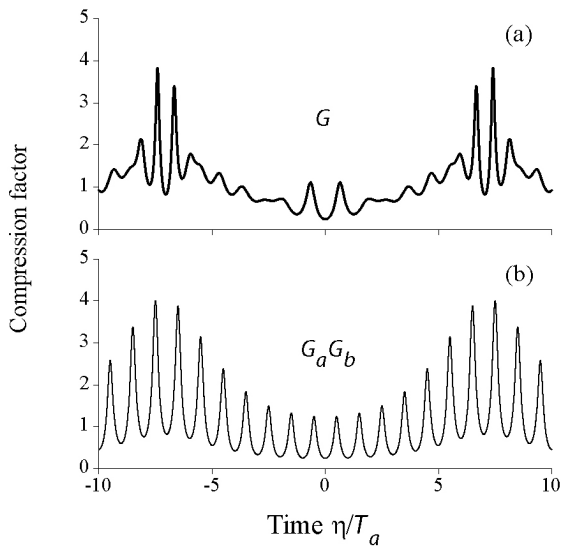}
 \end{center}
\caption{Comparison between the total  compression factor $G$ (a) and the product of 
individual compression factors $G_aG_b$ (b).
The plots are calculated  for the parameters $\alpha_a z=0.8$, $\alpha_b z=0.6$,  $\omega_b=0.07\,\omega_a$, and  $\varphi_a=\varphi_b=0$.
 }
\label{fig1}
\end{figure}

To see the interference between the individual susceptibility components 
in beating with the field,  
we illustrate the functions  $G$  and  $G_a G_b$ in Figs. \ref{fig1}(a) and \ref{fig1}(b),
respectively. The plots are calculated  for the parameters $\alpha_a z=0.8$, $\alpha_b z=0.6$, $\omega_b=0.07\,\omega_a$, and $\varphi_a=\varphi_b=0$. 
The time is normalized to the  Raman period $T_a=2\pi/\omega_a$. 
Both  $G$  and  $G_a G_b$ reveal  a sequence of teeth corresponding to the oscillation of  the   susceptibility component $\chi_a$, which has a higher frequency.  
The  peaks of  these teeth are modulated in  accordance with the oscillation of  the  component $\chi_b$,
which has a lower frequency.  The differences between  $G$  and  $G_a G_b$ are clearly observed.

\begin{figure}
\begin{center}
  \includegraphics{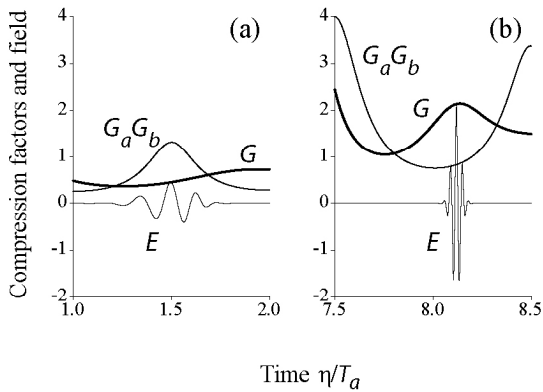}
 \end{center}
\caption{Magnification of the curves for  $G$ and $G_a G_b$ of Fig. \ref{fig1} in the regions
$1 \leq\eta/T_a\leq 2$ (a) and $7.5 \leq\eta/T_a\leq 8.5$ (b). 
The temporal profiles of a short probe field at the output in these time regions are also shown. 
The carrier frequency and  pulse length of the input  field  are $\omega_0=15.2\, \omega_a$ and 
$T=0.08\, T_a$,  respectively. The  input peak time is $\eta_p=0.83\, T_a$ (a) or $8.8\, T_a$ (b).
The  peak value of the input field is one unit.
}
\label{fig2}
\end{figure}

To see in detail   the difference between $G$ and $G_aG_b$,  we magnify in Fig. \ref{fig2} the curves of 
Fig. \ref{fig1} for  the time windows $1 \leq\eta/T_a\leq 2$ (a)  and $7.5 \leq\eta/T_a\leq 8.5$ (b). 
We also plot in Fig. \ref{fig2} the output field for the cases where the  input peak time is 
$\eta_p=0.83\, T_a$ (a) and $8.8\, T_a$ (b).  The carrier frequency and  pulse length of the input probe field  are $\omega_0=15.2\, \omega_a$ and 
$T=0.08\, T_a$,  respectively. 
We observe  $G<1<G_a G_b$ and $G>1>G_a G_b$ around the times $\eta/T_a=1.5$ and 8.1,  respectively. 
These relations imply that,   due to the interference between $\chi_a$ and $\chi_b$, 
the total compression factor $G$ and the product of individual factors $G_aG_b$ may give  opposite indications  
on pulse stretching and compression.
Thus, the interference between the susceptibility components  
can substantially affect frequency modulation and  pulse compression. 

In the above,  the total compression factor $G$ has been compared with the product $G_aG_b$ of the individual compression factors. However,  $G_aG_b$ 
cannot be rigorously considered as 
a compression factor for the case where there is no interference between $\chi_a$ and $\chi_b$.  
Below we will compare the compression factor  of a combined system
with the compression factor   of a sequence of two individual
systems.   

We call $M_aM_b$ a cascade   in which the probe field is sent through, first,  a cell $M_a$  (with susceptibility  $\chi_a$ and length $z_a$) 
and, then, a cell $M_b$  (with  susceptibility  $\chi_b$ and length $z_b$). For this configuration,  we have
$E _{\mathrm{out}}(\eta)= E _{\mathrm{in}}(s) G_{ab}(\eta)$ and $ds/d\eta=G_{ab}(\eta)$, where 
\begin{equation}
G_{ab}(\eta)=G_a(\eta', z_a)G_b(\eta, z_b).
\label{29}
\end{equation}
Here $\eta'$ and $s$ are determined by the equations \cite{Kalosha and Kien,Kien02}
\begin{eqnarray}
\tan[(\omega_b \eta'+\varphi_b)/2] =e^{-\alpha_b z_b}\tan[(\omega_b\eta+\varphi_b)/2], \nonumber\\   
\tan[(\omega_a s+\varphi_a)/2] =e^{-\alpha_a z_a}\tan[(\omega_a\eta'+\varphi_a)/2].     
\label{30}  
\end{eqnarray}
Note that $G_{ab}$ is not symmetric with respect to the indices  $a$ and $b$, i.e.,
$G_{ab}\not= G_{ba}$. 
In other words, the cascades $M_aM_b$ and $M_bM_a$ produce  different  compression factors.
This feature is  different from the 
results of Ref. \cite{HarrisSeries}, which are valid for the case of limited bandwidths.

\begin{figure}
\begin{center}
  \includegraphics{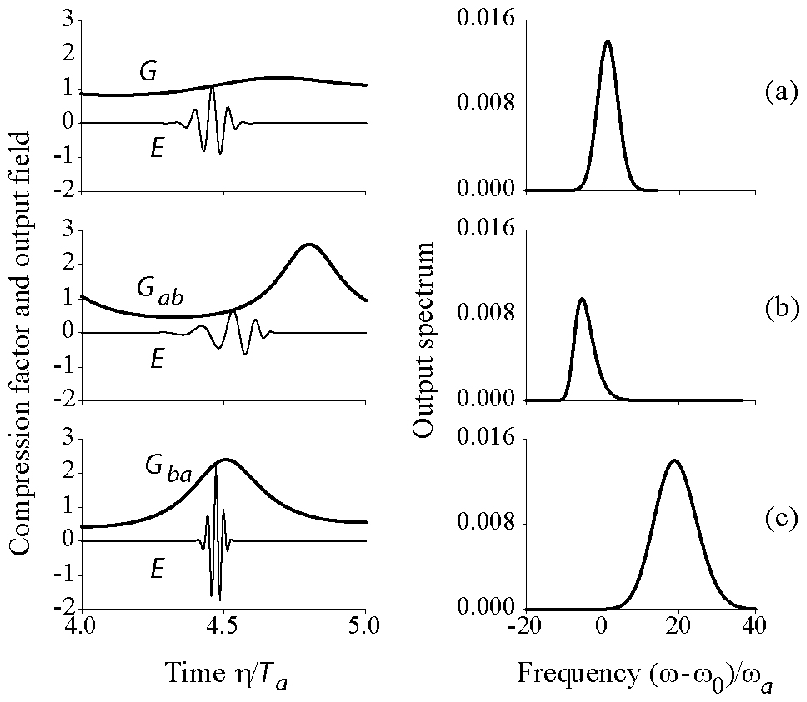}
 \end{center}
\caption{Comparison between  the mixture (a), the cascade $M_aM_b$ (b),   
and the  cascade $M_bM_a$ (c) of the  individual Raman  systems $M_a$ and $M_b$. 
The parameters  for the media  are $\alpha_a z=\alpha_a z_a=0.8$, $\alpha_b z=\alpha_b z_b=0.6$,  $\omega_b=0.07\,\omega_a$, and $\varphi_a=\varphi_b=0$.  
The carrier frequency and  pulse length of the input field  are the same as for Fig. \ref{fig2}.
The  input peak time is $\eta_p=3.1\, T_a$.
The  peak value of the input field is one unit.}
\label{fig3}
\end{figure}

Comparison between Eqs. (\ref{7}) and (\ref{29}) shows that the compression factor $G$ of the combined system
is different from the factors 
$G_{ab}$ and $G_{ba}$ of the cascades. 
To see  the difference between the mixture  and cascades of the individual systems $M_a$ and $M_b$,   
we plot the factors $G$, $G_{ab}$, and $G_{ba}$ in  Figs. \ref{fig3}(a),  \ref{fig3}(b), \ref{fig3}(c), respectively,
for  the time window $4 \leq\eta/T_a\leq 5$.
We also plot the output field and the output spectrum
for  the  input peak time  $\eta_p=3.1\, T_a$. 
The other parameters are the same as for Fig. \ref{fig2}. 
We observe $G_{ab}<G\cong1<G_{ba}$ around the time $\eta/T_a=4.5$.
These relations as well as the temporal and spectral profiles 
of the output field
show clearly  the differences between   the combined  and  cascade systems in frequency modulation and  pulse compression.

In the case where $\alpha_j z_j\ll1$ for $j=a,b$, 
we find to lowest order in $\alpha_j z_j$ the compression factor
$G_{ab}=1-\sum_j \alpha_j z_j\cos(\omega_j\eta+\varphi_j)$ and the 
input-output time relation
$s=\eta-\sum_j(\alpha_j z_j/\omega_j)\sin(\omega_j\eta+\varphi_j)$.  
Hence, with the use of the generating function 
$e^{i\xi\sin\theta}=\sum_{k=-\infty}^{\infty} J_k(\xi) e^{ik\theta}$
of the Bessel functions $J_k$, we can  expand an input oscillation $e^{i\omega_0s}$ into a series of
output harmonics $\exp[i(\omega_0+\sum_j q_j\omega_j)\eta]$.  
Then, we can approximate the spectrum of the field at the output of a series of  $M_a$ and $M_b$  
as  the product of the Bessel-function spectra of the individual  cells  \cite{HarrisSeries}. 
The condition $\alpha_jz_j\ll1$ means that  
the pulse bandwidths $\gamma_j z_j\omega_j$ \cite{Kien02,some theory}, 
with $\gamma_j=\alpha_j\omega_0/\omega_j$,
produced by  the individual cells, 
are small compared to the optical carrier frequency $\omega_0$.  
The same results are also obtained for the case of mixed cells
(except that individual cell lengths $z_j$ should be replaced by a common length $z$). 
Under the condition of small pulse bandwidths, there is no interference between $\chi_a$
and $\chi_b$, and therefore, no difference between the mixture
and cascades of $M_a$ and $M_b$ in  frequency modulation and  pulse compression.

\begin{figure}
\begin{center}
  \includegraphics{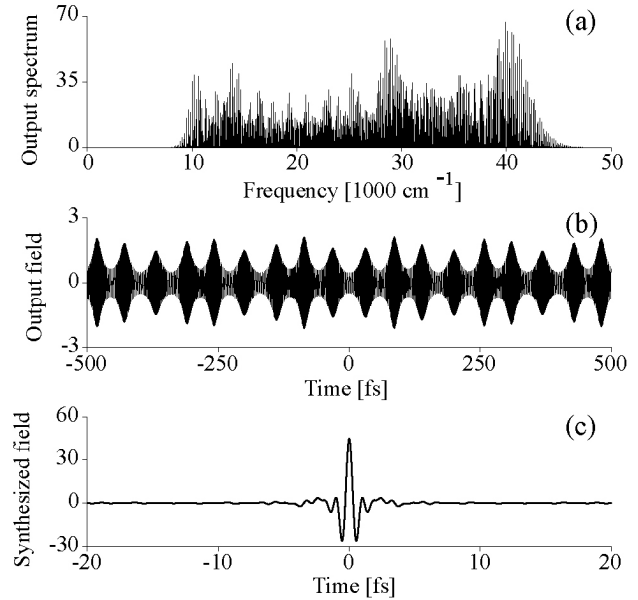}
 \end{center}
\caption{Spectrum (a), output field (b), and time-domain synthesis by a phase compensator (c)
for a cell containing  H$_2$ and D$_2$. 
The parameters  for the medium  are $\omega_a=587$ cm$^{-1}$, $\omega_b=179$ cm$^{-1}$,
$\alpha_a z=0.587$, $\alpha_b z=0.179$, and $\varphi_a=\varphi_b=0$.  
The carrier frequency,  pulse length, and peak time of the input field  are $\omega_0=20\,000$ cm$^{-1}$, $T=4.5$ ps, and $\eta_p=0$, respectively. 
The  peak value of the input field is one unit.
The synthesized field is obtained when the   sidebands are phase-corrected to the  same phase of zero.
}
\label{fig4}
\end{figure}

Finally, we demonstrate a numerical example for a realistic system, namely, for
a cell containing mixed H$_2$ and D$_2$
molecules. We take $\omega_a=587$ cm$^{-1}$ and $\omega_b=179$ cm$^{-1}$ so as to correspond to the rotational
transitions of H$_2$ and D$_2$. 
We use a probe with a carrier frequency $\omega_0=20\,000$ cm$^{-1}$, a pulse length  $T=4.5$ ps, and a peak time $\eta_p=0$. We assume the comb depths $\alpha_a z=0.587$ and $\alpha_b z=0.179$, which correspond to the
modulation depths $\gamma_a z=\gamma_b z=20$. 
Similar to the results of  Ref. \cite{HarrisSeries} for the case of a cascade,
a broad spectrum with a large  number of sidebands, ranging from 8000 cm$^{-1}$ to 46\,000 cm$^{-1}$,
is generated. Unlike the results of Ref. \cite{HarrisSeries}, the spectrum in Fig. \ref{fig4}(a) is asymmetric. 
Such asymmetry is due to the deviation  of the spectrum from the Bessel-function
spectrum, and occurs  when the comb depths $\alpha_j z$ are not too small compared to unity. 
The output field in Fig. \ref{fig4}(b)  is a long train of slightly compressed   sections.
As has been shown in Ref.  \cite{HarrisSeries} for the case of a cascade, 
single-cycle pulses can be synthesized when   the sidebands are phase-corrected by a phase compensator. 
We illustrate in Fig. \ref{fig4}(c) a single-cycle pulse synthesized from the phase-corrected
spectrum of Fig. \ref{fig4}(a).

In summary, we have studied  beating of a probe field with a  time-varying  susceptibility of a  Raman medium. We have derived a general analytical  solution and conservation relations  for this process. 
We have  shown that the interference between   Raman polarizations may  substantially 
affect frequency modulation and pulse compression.
We  emphasize that the analysis of this work can be directly 
applied only when dispersion is negligible.  When dispersion becomes 
substantial, our analytical solution and conservation relations still 
provide a useful insight, but a numerical simulation of pulse propagation 
is necessary in order to obtain an exact solution.

We thank A. K. Patnaik 
for helpful discussions. A. V. S. acknowledges support from the Welch Foundation.

\end{document}